\documentclass[sigconf,natbib=true,anonymous=false]{acmart}
\AtBeginDocument{%
  \providecommand\BibTeX{{%
    \normalfont B\kern-0.5em{\scshape i\kern-0.25em b}\kern-0.8em\TeX}}}

\usepackage{booktabs} 

\usepackage[english]{babel}
\usepackage{moresize}
\usepackage{amsmath}
\usepackage[ruled,linesnumbered,vlined]{algorithm2e}
\usepackage{balance}
\usepackage{comment}
\usepackage{paralist}
\usepackage{bm}
\usepackage{pgfplots}
\usetikzlibrary{pgfplots.dateplot}

\usepackage{subfigure}
\usepackage{flushend}
\usepackage[english]{babel}
\usepackage[latin1]{inputenc}
\usepackage{mathrsfs}
\usepackage{graphicx}

\usepackage{amssymb}
\usepackage{amsfonts}
\usepackage{url}
\usepackage{longtable}
\usepackage{rotating}
\usepackage{multirow}
\usepackage{mathrsfs}
\usepackage{enumitem}
\usepackage{adjustbox}
\usepackage{hyperref}
\usepackage{pgfplots}
\usetikzlibrary{pgfplots.dateplot}
\usepackage{filecontents}
\usepackage{threeparttable}
\definecolor{tblue}{RGB}{31,119,180}
\definecolor{torange}{RGB}{255,127,14}
\definecolor{tgreen}{RGB}{44,160,44}
\definecolor{tred}{RGB}{214,39,40}
\definecolor{tpurple}{RGB}{148,103,189}

\newcommand{\hide}[1]{} 

\def\model{DiffKG\xspace}

\begin{document}

\title{DiffKG: Knowledge Graph Diffusion Model for Recommendation}


\author{Yangqin Jiang}
\affiliation{%
  \institution{University of Hong Kong}
  \city{Hong Kong}
  \country{China}
}
\email{mrjiangyq99@gmail.com}

\author{Yuhao Yang}
\affiliation{%
  \institution{University of Hong Kong}
  \city{Hong Kong}
  \country{China}
}
\email{yuhao-yang@outlook.com}

\author{Lianghao Xia}
\affiliation{%
  \institution{University of Hong Kong}
  \city{Hong Kong}
  \country{China}
}
\email{aka_xia@foxmail.com}

\author{Chao Huang}
\authornote{Chao Huang is the corresponding author.}
\affiliation{%
  \institution{University of Hong Kong}
  \city{Hong Kong}
  \country{China}
}
\email{chaohuang75@gmail.com}

\renewcommand{\shortauthors}{Yangqin Jiang, Yuhao Yang, Lianghao Xia, \& Chao Huang}

\begin{abstract}
Knowledge Graphs (KGs) have emerged as invaluable resources for enriching recommendation systems by providing a wealth of factual information and capturing semantic relationships among items. Leveraging KGs can significantly enhance recommendation performance. However, not all relations within a KG are equally relevant or beneficial for the target recommendation task. In fact, certain item-entity connections may introduce noise or lack informative value, thus potentially misleading our understanding of user preferences. To bridge this research gap, we propose a novel knowledge graph diffusion model for recommendation, referred to as \model. Our framework integrates a generative diffusion model with a data augmentation paradigm, enabling robust knowledge graph representation learning. This integration facilitates a better alignment between knowledge-aware item semantics and collaborative relation modeling. Moreover, we introduce a collaborative knowledge graph convolution mechanism that incorporates collaborative signals reflecting user-item interaction patterns, guiding the knowledge graph diffusion process. We conduct extensive experiments on three publicly available datasets, consistently demonstrating the superiority of our \model compared to various competitive baselines. We provide the source code repository of our proposed \model model at the following link: \color{blue}\url{https://github.com/HKUDS/DiffKG}.

\end{abstract}

\copyrightyear{2024}
\acmYear{2024}
\setcopyright{acmlicensed}\acmConference[WSDM '24]{Proceedings of the 17th
ACM International Conference on Web Search and Data Mining}{March 4--8,
2024}{Merida, Mexico}
\acmBooktitle{Proceedings of the 17th ACM International Conference on Web
Search and Data Mining (WSDM '24), March 4--8, 2024, Merida, Mexico}
\acmDOI{10.1145/3616855.3635850}
\acmISBN{979-8-4007-0371-3/24/03}



\maketitle
\section{Introduction}
\label{sec:intro}

In the context of the information overload problem, recommendation systems have gained substantial influence in the modern web landscape. These systems have become an integral part of the online experience by effectively connecting users with items that align with their individual interests. Collaborative filtering (CF), one of the leading paradigms for recommendation systems, postulates that users who engage in similar interaction modes also share similar interests towards items. This approach has garnered considerable attention and proven to be highly effective in delivering personalized recommendations to users~\cite{koren2021advances, liang2018variational, ren2023disentangled,rendle2012bpr}.

The recommendation performance in practical scenarios is significantly hindered by the inherent sparsity of user-item interactions~\cite{yao2021self,wei2023llmrec}. To mitigate this issue, the integration of a knowledge graph (KG) as a comprehensive information network for items has emerged as a new trend in collaborative filtering, known as knowledge-aware recommendation. Researchers have explored knowledge-aware recommendation through two primary approaches: embedding-based methods and path-based methods. Embedding-based methods~\cite{cao2019unifying, wang2018dkn, zhang2016collaborative} have been employed to enhance the modeling of users and items by incorporating transition-based knowledge graph embeddings into item representations. On the other hand, path-based methods~\cite{wang2019explainable, yu2014personalized} focus on extracting semantically meaningful meta-paths from the knowledge graph and leveraging them to perform complex modeling of users and items. To combine the strengths of embedding-based and path-based methods, recent research has turned to GNNs as a powerful tool. GNN methods leverage the capabilities of propagation and aggregation over the knowledge graph to capture high-order information~\cite{wang2019knowledge,wang2019kgat,wang2021learning}.

Despite the demonstrated effectiveness of existing knowledge graph (KG)-aware recommendation methods, their performance heavily relies on high-quality input knowledge graphs and can be adversely affected by the presence of noise. In practical scenarios, knowledge graphs often suffer from sparsity and noise, characterized by long-tail entity distributions and topic-irrelevant connections between items and entities~\cite{pujara2017sparsity, wang2018label}. To address these challenges, recent research has proposed the utilization of contrastive learning (CL) techniques to enhance knowledge-aware recommendation. For instance, the KGCL approach~\cite{yang2022knowledge} leverages stochastic graph augmentation on the knowledge graph and employs CL to address the long-tail issues within the KG. Likewise, the MCCLK~\cite{zou2022multi} and KGIC~\cite{zou2022improving} methods introduce cross-view CL paradigms between the knowledge graph and user-item graph, aiming to integrate external item knowledge into the modeling of user-item interactions. However, it is worth noting that these methods predominantly rely on simplistic random augmentation or intuitive cross-view information, overlooking the substantial amount of irrelevant information present in the knowledge graph for the specific recommendation task. Thus, it is of paramount importance to effectively filter out noisy knowledge graph information, leading to a more resilient encoding of user preferences.

This research introduces an innovative model known as \model for knowledge-aware recommender systems. Drawing inspiration from recent advancements in diffusion models, we propose a unique knowledge graph diffusion paradigm that effectively balances corruption and reconstruction. Our approach involves a progressive forward process where the initial knowledge graph undergoes step-by-step corruption through the introduction of random noises. This incremental corruption process accumulates noises over multiple iterations, which are then iteratively recovered to restore the original knowledge graph structures. By employing this tractable forward process, we establish a feasible posterior and enable reverse generation using flexible neural networks to model complex distributions iteratively. To address the challenge of noisy information within the knowledge graph, we introduce a KG filter that eliminates irrelevant and erroneous data, aligning seamlessly with the learning of user preferences. Additionally, we devise a collaborative knowledge graph convolution mechanism, which enhances our diffusion model by integrating collaborative signals into the KG diffusion process. It ensures the retention of relevant knowledge during the diffusion process. Furthermore, we propose a KG diffusion-enhanced data augmentation paradigm to benefit the model with the enriched information and improved learning capabilities.

In summary, this paper makes the following contributions: \vspace{-0.05in}
\begin{itemize}[leftmargin=*]
    \item We present a novel recommendation model called \model, which leverages task-relevant item knowledge to enhance the collaborative filtering paradigm. Our approach introduces a new framework that allows for the distillation of high-quality signals from the aggregated representation of noisy knowledge graphs. \\\vspace{-0.12in}
    \item We propose an integration of the generative diffusion model with the knowledge graph learning framework, designed for knowledge-aware recommendation. This integration allows us to effectively align the semantics of knowledge-aware items with collaborative relation modeling for recommendation purposes. \\\vspace{-0.12in}
    \item Our extensive experimental evaluations substantiate the substantial performance gains achieved by our \model framework when compared to various baseline models across diverse benchmark datasets. Notably, our approach effectively tackles the challenges stemming from data noise and data scarcity, which are known to exert a negative impact on the accuracy of recommendation.

\end{itemize}
\section{Preliminaries}
\label{sec:model}
We introduce the key concepts that form the paper foundation and provide a formal definition of the KG-enhanced recommendation. \\\vspace{-0.12in}

\noindent \textbf{User-Item Interaction Graph.}
Consider a typical recommendation scenario with a set of users denoted as $\mathcal{U}$ and a set of items denoted as $\mathcal{I}$. Each individual user $u$ belongs to the set $\mathcal{U}$, and each item $i$ belongs to the set $\mathcal{I}$. To represent the collaborative signals between users and items, we construct a binary graph denoted as $\mathcal{G}_u = {(u, y_{u,i}, i)}$. Here, $y_{ui} = 1$ indicates that user $u$ has interacted with item $i$, while $y_{u,i}=0$ signifies the absence of such interaction. \\\vspace{-0.12in}

\noindent \textbf{Knowledge Graph.}
The knowledge graph is denoted as $\mathcal{G}_{k} = {(h, r, t)}$ and serves to organize external item attributes by incorporating various types of entities and their corresponding relationships. Each triplet ($h$, $r$, $t$) within the knowledge graph characterizes the semantic relatedness between the head entity $h$ and the tail entity $t$, connected by the relation $r$. The entities $h$ and $t$ encompass items and their associated concepts, such as directors for movies. By utilizing this supplementary knowledge graph, we can effectively model and analyze the intricate relationships that exist between items and entities. This, in turn, empowers us to gain a more comprehensive and nuanced understanding of the item attributes. \\\vspace{-0.12in}

We define the KG-enhanced recommendation task as follows: given the user-item interaction graph $\mathcal{G}_u$ and the associated knowledge graph $\mathcal{G}_k$, our objective is to train a recommender model $\mathcal{F}(u,i|\mathcal{G}_u, \mathcal{G}_k, \Theta)$ with learnable parameters $\Theta$. This model aims to predict the likelihood of user $u$ interacting with item $i$. 

\section{The Proposed \model Framework}
\label{sec:solution}
\begin{figure}[t]
    \centering
    \includegraphics[width=0.85 \linewidth]{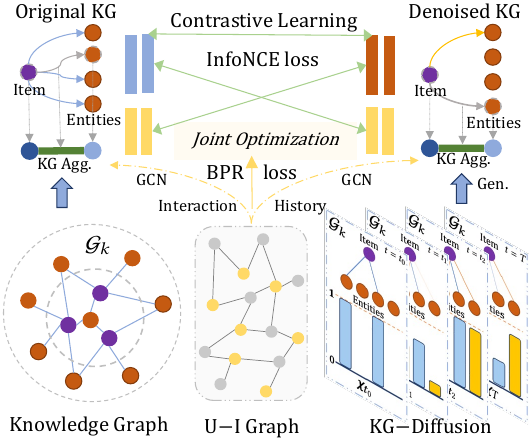}
    \caption{Overall framework of the proposed \model model.}
    \label{fig:figure_overall}
    \vspace{-0.1in}
\end{figure}

In this section, we present the technical design of our proposed \model, accompanied by the overall model architecture depicted in Fig.~\ref{fig:figure_overall}. Our model includes a heterogeneous knowledge aggregation module, a knowledge graph diffusion model, and a KG diffusion-enhanced data augmentation paradigm. These components effectively capture diverse relationships in the KG and ensure high-quality KG information for enhancing recommendation. 

\subsection{Heterogeneous Knowledge Aggregation}
To handle the heterogeneity of knowledge relations in real-world knowledge graphs, we employ a relation-aware knowledge embedding layer inspired by graph attention mechanisms utilized in previous works such as~\cite{velivckovic2017graph, wang2019kgat, yang2022knowledge}. This layer enables effective capturing of diverse relationships inherent in the connection structure of the knowledge graph. By incorporating a parameterized attention matrix, it projects entity-dependent context and relation-dependent context into specific representations, overcoming the limitations of manually designing path generation on knowledge graphs. The message aggregation mechanism between an item and its connected entities can be described as follows:
\begin{equation}
\begin{split}
    \mathbf{x}_i = &Drop(Norm(\mathbf{x}_i + \sum_{e\in\mathcal{N}_i}\alpha(e, r_{e,i}, i) \mathbf{x}_e)),\\
    \alpha(e, r_{e,i}) &= \frac{\text{exp}(LeakyReLU(r_{e,i}^{T}W[\mathbf{x}_e||\mathbf{x}_i]))}{\sum_{e\in\mathcal{N}_i}\text{exp}(LeakyReLU(r_{e,i}^{T}W[\mathbf{x}_e||\mathbf{x}_i])))}
\end{split}
\end{equation}

In the knowledge aggregation process, $\mathcal{N}_i$ represents the neighboring entities of item $i$ based on different types of relations $r_{e,i}$ in the knowledge graph $\mathcal{G}_k$. The embeddings of the item and entity are denoted as $\mathbf{x}_i \in \mathbb{R}^d$ and $\mathbf{x}_e \in \mathbb{R}^d$, respectively. To prevent overfitting, we apply the dropout function denoted as $Drop$, and for normalization, we use the function $Norm$. The term $\alpha (e, r_{e,i}, i)$ represents the estimated entity-specific and relation-specific attentive relevance during the knowledge aggregation process, capturing the distinct semantics of relationships between item $i$ and entity $e$. A parametric weight matrix $W \in \mathbb{R}^{d\times2d}$ is customized to the input item and entity representations, and a non-linear transformation is induced using the $LeakyReLU$ activation function. Notably, we incorporate random dropout operations on the knowledge graphs before heterogeneous knowledge aggregation. This is because a sparse knowledge graph inherently has the potential to significantly enhance the performance of the recommender system.

\subsection{KG-enhanced Data Augmentation}

Contrastive learning has recently gained remarkable success in the realm of recommendation systems. In the context of knowledge graph-enhanced recommendation, methods like KGCL~\cite{yang2022knowledge}, MCCLK~\cite{zou2022multi}, and KGIC~\cite{zou2022improving} have introduced contrastive learning techniques. However, these approaches often rely on simplistic random augmentation methods or simplistic cross-view contrasts between the raw knowledge graph view and collaborative filtering view. Unfortunately, the random augmentation can introduce unwanted noise, and the supplementary knowledge graph view may contain irrelevant information. It is crucial to acknowledge that within the wealth of semantic relationships present in a knowledge graph, only a subset is truly relevant to the downstream recommendation task. Failing to address these irrelevant knowledge relationships can have a detrimental impact on the recommendation performance.

To tackle these challenges, we propose the use of a generative model to reconstruct a subgraph $\mathcal{G}_k^{'}$ of the knowledge graph $\mathcal{G}_k$ that specifically contains the relationships relevant to the downstream recommendation task. In Section~\ref{subsec:kgdm}, we will provide a detailed explanation of this generative model. Once we have constructed the task-related knowledge graph, we encode the representations of users and items using a combination of the graph-based collaborative filtering framework and heterogeneous knowledge aggregation. Taking inspiration from the effectiveness of the simplified graph convolutional network in LightGCN~\cite{he2020lightgcn}, we design our own local graph embedding propagation layer, which can be described as:
\begin{equation}
    \mathbf{x}_u^{(l+1)} = \sum_{i\in\mathcal{N}_u}\frac{\mathbf{x}_i^{(l)}}{\sqrt{|\mathcal{N}_{u}|\cdot |\mathcal{N}_{i}|}};\mathbf{x}_i^{(l+1)} = \sum_{u\in\mathcal{N}_i}\frac{\mathbf{x}_u^{(l)}}{\sqrt{|\mathcal{N}_{i}|\cdot |\mathcal{N}_{u}|}},
\end{equation}
We utilize $\mathbf{x}_u^{(l)}$ and $\mathbf{x}_i^{(l)}$ to represent the encoded representations of user $u$ and item $i$ at the $l$-th graph propagation layer. The neighboring items/users of user $u$/item $i$ are denoted as $\mathcal{N}_u$ and $\mathcal{N}_i$ respectively. By employing multiple graph propagation layers, the graph-based collaborative filtering (CF) framework captures collaborative signals of higher order. In our encoding pipeline, both $\mathcal{G}_k$ and $\mathcal{G}_k^{'}$ are employed for heterogeneous knowledge aggregation, allowing us to generate input item feature vectors while preserving the semantic information of the knowledge graph. These item embeddings are subsequently fed into the graph-based CF framework to refine their representations further.

Once we have established two knowledge-enhanced graph views, we consider the view-specific embeddings of the same node as positive pairs (e.g., {($\mathbf{x}_{u}^{'}$, $\mathbf{x}_{u}^{''}$)|$u \in \mathcal{U}$}). On the other hand, we regard the embeddings of different nodes in the two views as negative pairs (e.g., {($\mathbf{x}_{u}^{'}$, $\mathbf{x}_{v}^{''}$)|$u,v \in \mathcal{U}$, $u \neq v$}). To formalize this, we define a contrastive loss function that aims to maximize the agreement among positive pairs and minimize the agreement among negative pairs. The contrastive loss can be expressed as follows:
\begin{equation}
    \label{eq:infoNCE}
    \mathcal{L}_{cl}^{user} = \sum_{u\in\mathcal{U}} -\text{log} \frac{\text{exp}(s(\mathbf{x}_{u}^{'}, \mathbf{x}_{u}^{''})/\tau)}{\sum_{v \in \mathcal{U}}\text{exp}(s(\mathbf{x}_{u}^{'},\mathbf{x}_{v}^{''})/\tau)},
\end{equation}
The similarity between two vectors is measured using the cosine similarity function, denoted as $s(\cdot)$. The hyper-parameter $\tau$, referred to as the temperature, is used in the softmax operation. We obtain the contrastive loss of the user side as $\mathcal{L}_{cl}^{user}$, and similarly, we compute the contrastive loss of the item side as $\mathcal{L}_{cl}^{item}$. By combining these two losses, we obtain the objective function for the self-supervised task, which can be represented as $\mathcal{L}{cl} = \mathcal{L}_{cl}^{user} + \mathcal{L}_{cl}^{item}$.

\subsection{Diffusion with Knowledge Graph}
\label{subsec:kgdm}
Drawing inspiration from the effectiveness of diffusion models in data generation from noisy inputs, such as diffusion models presented in works like~\cite{wang2023diffusion, ho2020denoising, sohl2015deep}, we propose a knowledge graph diffusion model. Our purpose is to generate a recommendation-relevant subgraph $\mathcal{G}_k^{'}$ from the original knowledge graph $\mathcal{G}_k$. To achieve this, the model is trained to identify true relationships between items and entities in a knowledge graph that has been corrupted by a noise diffusion process. Our method employs a forward process that gradually introduces noise to the relations in the knowledge graph, simulating the corruption of relations. Then, through iterative learning, we aim to recover the original relations in the knowledge graph. This iterative denoising training enables \model to model complex relation generation procedures and reduce the impact of noisy relations. Ultimately, the restored relation probabilities are utilized to reconstruct the subgraph $\mathcal{G}_k^{'}$ from the original knowledge graph $\mathcal{G}_k$.

\subsubsection{\bf Noise Diffusion Process.}
\begin{figure}[t]
    \centering
    \includegraphics[width=0.85 \linewidth]{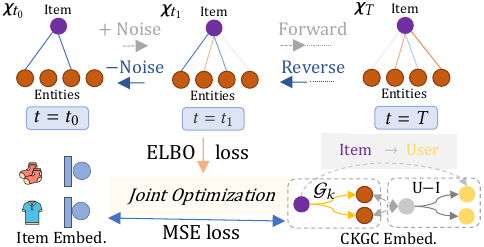}
    \caption{Diffusion Model with Knowledge Graph.}
    \vspace{-0.1in}
    \label{fig:figure_KGDM}
\end{figure}

In Fig.~\ref{fig:figure_KGDM}, we can observe that our knowledge graph (KG) diffusion, similar to other diffusion models, consists of two essential processes: the forward process and the reverse process. In order to apply these processes to the KG, we represent the KG using an adjacency matrix. Specifically, let's consider an item $i$ that has relations with entities in the entity set $\mathcal{E}$. We denote these relations as $\mathbf{z}_i = [\mathbf{z}_i^{0}, \mathbf{z}_i^{1}, \cdots, \mathbf{z}_i^{|\mathcal{E}|-1}]$, where $\mathbf{z}_i^{e} = 1$ or $0$. This binary value indicates whether item $i$ has a relation with entity $e$ or not. In the forward process, the original structure of the knowledge graph (KG) is corrupted by adding Gaussian noises step by step. We initialize the initial state $\bm{\chi}_{0}$ as the original adjacency matrix $\mathbf{z}_i$ of the item. This means that $\bm{\chi}_{0} = \mathbf{z}_i$. The forward process then constructs $\bm{\chi}_{1:\textit{T}}$ in a Markov chain by gradually adding Gaussian noise in $\textit{T}$ steps. We parameterize the transition from $\bm{\chi}_{t-1}$ to $\bm{\chi}_{t}$ as:
\begin{equation}
\label{eq:q_x_t}
    q(\bm{\chi}_{t}|\bm{\chi}_{t-1}) = \mathcal{N}(\bm{\chi}_{t};\sqrt{1-\beta_t}\bm{\chi}_{t-1}, \beta_t\textbf{\emph{I}}),
\end{equation}
$t \in {1, \cdots, \textit{T}}$ represents the diffusion step. $\mathcal{N}$ denotes the Gaussian distribution, and $\beta_t \in (0, 1)$ controls the scale of the Gaussian noise added at each step $t$. As $\textit{T} \rightarrow \infty$, the state $\bm{\chi}_{\textit{T}}$ converges towards a standard Gaussian distribution. By utilizing the reparameterization trick and taking advantage of the additivity property of two independent Gaussian noises, we can directly derive the state $\bm{\chi}_{t}$ from the initial state $\bm{\chi}_{0}$. Formally, we describe this process as follows:
\begin{equation}
    \label{eq:q_eq}
    q(\bm{\chi}_{t}|\bm{\chi}_{0}) = \mathcal{N}(\bm{\chi}_{t};\sqrt{\bar{\alpha}_t}\bm{\chi}_{0}, (1 - \bar{\alpha}_t)\textbf{\emph{I}}), \bar{\alpha}_t = \prod_{t^{'}=1}^t (1 - \beta_{t^{'}}).
\end{equation}
$\bm{\chi}_{t}$ can be reparameterized as follows:
\begin{equation}
    \bm{\chi}_{t} = \sqrt{\bar{\alpha}_t} \bm{\chi}_{0} + \sqrt{1-\bar{\alpha}_t}\bm{\epsilon}, \bm{\epsilon} \sim \mathcal{N}(0, \textbf{\emph{I}}).
\end{equation}
To regulate the addition of noises in $\bm{\chi}_{1:\textit{T}}$, we incorporate a linear noise scheduler that implements $1 - \bar{\alpha}t$ using three hyperparameters: $s$, $\alpha{low}$, and $\alpha{up}$. The linear noise scheduler is defined as follows:
\begin{equation}
    1 - \bar{\alpha}_t = s \cdot \left[\alpha_{low} + \frac{t-1}{\textit{T}-1}(\alpha_{up} - \alpha_{low})\right], t \in \{1,\cdots,\textit{T}\}.    
\end{equation}
The linear noise scheduler uses three hyperparameters: $s \in [0, 1]$ controls the noise scales, while $\alpha_{low} < \alpha_{up} \in (0, 1)$ set the upper and lower bounds for the added noises.

Next, the diffusion model learns to remove the added noises from $\bm{\chi}_{t}$ in order to recover $\bm{\chi}_{t-1}$ using neural networks. Starting from $\bm{\chi}_{\textit{T}}$, the reverse process gradually reconstructs the relations within the knowledge graph (KG) through the denoising transition step. The denoising transition step is outlined as follows:
\begin{equation}
    p_{\theta}(\bm{\chi}_{t-1}|\bm{\chi}_{t}) = \mathcal{N}(\bm{\chi}_{t-1};\bm{\mu}_\theta(\bm{\chi}_{t},t),\bm{\Sigma}_\theta(\bm{\chi}_{t},t)).
\end{equation}
We utilize neural networks parameterized by $\theta$ to generate the mean $\bm{\mu}\theta(\bm{\chi}_{t},t)$ and covariance $\bm{\Sigma}\theta(\bm{\chi}_{t},t)$ of a Gaussian distribution. 

\subsubsection{\bf Optimization of KG Diffusion Process.}
To optimize our model, we maximize the Evidence Lower Bound (ELBO) of the likelihood of the original knowledge graph relations $\bm{\chi}_{0}$. Following the approach described in \cite{wang2023diffusion}, we can summarize the optimization objective of our probabilistic diffusion process as follows:
\begin{equation}
\begin{split}
\label{eq:diffTrain}
     \text{log}p(\bm{\chi}_{0}) &\ge \mathbb{E}_{q(\bm{\chi}_{1}|\bm{\chi}_{0})}[\text{log}p_\theta(\bm{\chi}_{0}|\bm{\chi}_{1})]\\
     - &\sum_{t=2}^{\textit{T}}\mathbb{E}_{q(\bm{\chi}_{t}|\bm{\chi}_{0})}[\textit{D}_{KL}(q(\bm{\chi}_{t-1}|\bm{\chi}_{t},\bm{\chi}_{0})||p_\theta(\bm{\chi}_{t-1}|\bm{\chi}_{t}))].
\end{split}
\end{equation}
The optimization objective of diffusion model consists of two terms. The first term measures the recovery probability of $\bm{\chi}_{0}$, representing the ability of the model to reconstruct the original knowledge graph. The second term regulates the recovery of $\bm{\chi}_{t-1}$ for $t$ ranging from $2$ to $\textit{T}$ in the reverse process.

The second term in the optimization objective aims to make the distribution $p_\theta(\bm{\chi}_{t-1}|\bm{\chi}_{t})$ approximate the tractable distribution $q(\bm{\chi}_{t-1}|\bm{\chi}_{t}, \bm{\chi}_{0})$ through the KL divergence $\textit{D}_{KL}(\cdot)$.
Following \cite{wang2023diffusion}, the second term $\mathcal{L}_t$ at step $t$ is as follows:
\begin{equation}
\label{eq:denoiseFinalLoss}
    \mathcal{L}_t=\mathbb{E}_{q(\bm{\chi}_{t}|\bm{\chi}_{0})}\left[\frac{1}{2}\left(\frac{\bar{\alpha}_{t-1}}{1-\bar{\alpha}_{t-1}}-\frac{\bar{\alpha}_t}{1-\bar{\alpha}_t}\right)||\hat{\bm{\chi}}_{\theta}(\bm{\chi}_{t},t)-\bm{\chi}_{0}||_{2}^{2}\right],
\end{equation}
where $\hat{\bm{\chi}}_{\theta}(\bm{\chi}_{t}, t)$ is the predicted $\bm{\chi}_0$ based on $\bm{\chi}_t$ and $t$.
To calculate Eq.~\ref{eq:denoiseFinalLoss}, we implement $\hat{\bm{\chi}}_{\theta}(\bm{\chi}_{t},t)$ by neural networks.
Specifically, we instantiate $\hat{\bm{\chi}}_{\theta}(\cdot)$ via a Multi-Layer Perceptron (MLP) that takes $\bm{\chi}_{t}$ and the step embedding of $t$ as inputs to predict $\bm{\chi}_{0}$.

For the first term, we use $\mathcal{L}_{first}$ to denote the negative of the first term in Eq.~\ref{eq:diffTrain} and it can be calculate as follows:
\begin{equation}
\begin{split}
\label{eq:reconstructFinalLoss}
    \mathcal{L}_{first}&\triangleq-\mathbb{E}_{q(\bm{\chi}_{1}|\bm{\chi}_{0})}[\text{log}p_\theta(\bm{\chi}_{0}|\bm{\chi}_{1})]\\
    &=\mathbb{E}_{q(\bm{\chi}_{1}|\bm{\chi}_{0})}\left[||\hat{\bm{\chi}}_{\theta}(\bm{\chi}_{1},1)-\bm{\chi}_{0}||_{2}^{2}\right],
\end{split}
\end{equation}
where we estimate the Gaussian log-likelihood log $p(\bm{\chi}_{0}|\bm{\chi}_{1})$ by unweighted $-||\hat{\bm{\chi}}_{\theta}(\bm{\chi}_{1},1)-\bm{\chi}_{0}||_{2}^{2}$.
It is easy to find that $\mathcal{L}_{first}$ is equal to $\mathcal{L}_1$ based on Eq.~\ref{eq:denoiseFinalLoss}.
Therefore, the first term in Eq.~\ref{eq:diffTrain} can be considered as $-\mathcal{L}_{1}$.

According to Eq.~\ref{eq:denoiseFinalLoss}, ELBO in Eq.~\ref{eq:diffTrain} can be formulated as $-\mathcal{L}_1-\sum_{t=2}^{\textit{T}}\mathcal{L}_t$.
Hence, to maximize the ELBO, we can optimize $\theta$ in $\hat{\bm{\chi}}_{\theta}(\bm{\chi}_{t},t)$ by minimizing $\sum_{t=1}^{\textit{T}}\mathcal{L}_t$.
Specifically, we uniformly sample step $t$ to optimize $\mathcal{L}_{elbo}$ over $t \sim \mathcal{U}(1, \textit{T})$. 
Formally, the ELBO loss $\mathcal{L}_{elbo}$ is shown below:
\begin{equation}
    \mathcal{L}_{elbo} = \mathbb{E}_{t\sim\mathcal{U}(1,\textit{T})}\mathcal{L}_t.
\end{equation}

\subsubsection{\bf Knowledge Graph Generation with Diffusion Model.}
In contrast to other diffusion models that randomly draw Gaussian noises for reverse generation, we have designed a simple inference strategy that aligns with the training of \model for relation prediction in knowledge graphs (KGs). This strategy avoids corrupting the KG with pure noises, as doing so would severely compromise the informative structure of the KG.

In our inference strategy, we begin by corrupting the original KG relations $\bm{\chi}_{0}$ in a step-by-step manner during the forward process, resulting in $\bm{\chi}_{\textit{T}^{'}}$. We then set $\hat{\bm{\chi}}_{\textit{T}} = \bm{\chi}_{\textit{T}^{'}}$ and perform reverse denoising, where we ignore the variance and use $\hat{\bm{\chi}}_{t-1} = \mu\theta(\hat{\bm{\chi}}_{t}, t)$ for deterministic inference. Next, we reconstruct the structure of the modified KG $\mathcal{G}_{k}^{'}$ using $\hat{\bm{\chi}}_{0}$. For each item $i$, we select the top $k$ $\hat{\mathbf{z}}_i^{j}$ ($j \in [0,|\mathcal{E}|-1]$, $j \in \mathcal{J}$, and $|\mathcal{J}|=k$) and add $k$ relations between item $i$ and entities $j \in \mathcal{J}$. It aims to preserve the informative structure of the KG while incorporating noise during the forward process and deterministic inference during the reverse process.

\subsubsection{\bf Collaborative Knowledge Graph Convolution.}
To mitigate the potential limitations of the diffusion model in generating a denoised knowledge graph that encompasses pertinent relationships for downstream recommendation tasks, we propose a collaborative knowledge graph convolution (CKGC) mechanism. This novel approach capitalizes on the user-item interaction data to assimilate supervisory signals from recommendation tasks into the optimization of KG diffusion. Through the aggregation of user-item interaction data, our method enhances the model's capacity to capture user preferences and seamlessly incorporates them into the denoised knowledge graph, thereby enhancing its relevance to recommendation tasks. This amalgamation of user preferences introduces a valuable dimension to the optimization process of KG diffusion, effectively bridging the divide between knowledge graph denoising and recommendation tasks.

The loss of collaborative knowledge graph convolution, denoted as $\mathcal{L}_{ckgc}$, is computed by incorporating user-item interaction information and knowledge graph predictions into the item embedding generation process. Specifically, we begin by aggregating the user-item interaction information $\mathcal{A}$ with the predicted relation probabilities from the knowledge graph, represented as $\hat{\bm{\chi}}_{0}$. This aggregation updates the user-item interaction matrix, effectively integrating the knowledge graph information. Next, we combine this updated user-item matrix with the user embeddings $\mathbf{E}_{u}$ to obtain an item embedding $\mathbf{E}_{i}^{'}$ that jointly incorporates both the knowledge graph and user information. Finally, we calculate the mean squared error (MSE) loss between the aggregated item embedding $\mathbf{E}_{i}^{'}$ and the original item embedding $\mathbf{E}_{i}$, and optimize it alongside the ELBO loss ($\mathcal{L}_{elbo}$). The formal expression for the loss $\mathcal{L}_{ckgc}$ is as follows: 
\begin{equation}
    \mathcal{L}_{ckgc} = \left\|\left[\mathcal{A}\cdot\hat{\bm{\chi}}_{0}^{\top}\right]^{\top}\cdot\mathbf{E}_{u}-\mathbf{E}_{i}\right\|_{2}^{2}
\end{equation}

\subsection{The Learning Process of \model}
The training of our \model consists of two primary components: training for the recommendation task and training for KG diffusion. The joint training of KG diffusion encompasses two loss components: the ELBO loss and the CKGC loss, which are optimized simultaneously. As a result, the loss function for KG diffusion can be expressed as follows:
\begin{equation}
    \mathcal{L}_{kgdm} = (1-\lambda_0)\mathcal{L}_{elbo} + \lambda_0\mathcal{L}_{ckgc},
\end{equation}
To balance the contributions of the ELBO loss and the CKGC loss, we introduce a hyperparameter $\lambda_0$ that controls their respective strengths. For the recommendation task, we incorporate the original Bayesian personalized ranking (BPR) recommendation loss along with the contrastive loss $\mathcal{L}_{cl}$ mentioned earlier. The BPR loss, denoted as $\mathcal{L}_{bpr}$, is defined as follows:
\begin{equation}
    \label{eq:bpr}
    \mathcal{L}_{bpr} = \sum_{(u,i,j)\in\mathcal{O}} - \text{log}\sigma(\hat{y}_{ui}-\hat{y}_{uj}),
\end{equation}
The training data is represented as $\mathcal{O} = {(u,i,j)|(u,i) \in \mathcal{O}^{+}, (u,j) \in \mathcal{O}^{-}}$, where $\mathcal{O}^{+}$ denotes the observed interactions and $\mathcal{O}^{-}$ represents the unobserved interactions obtained from the Cartesian product of user set $\mathcal{U}$ and item set $\mathcal{I}$ excluding $\mathcal{O}^{+}$. With these definitions, the integrative optimization loss for the recommendation task is:
\begin{equation}
    \mathcal{L}_{rec} = \mathcal{L}_{bpr} + \lambda_1\mathcal{L}_{cl} + \lambda_2||\theta||_{2}^{2},
\end{equation}
The learnable model parameters are denoted as $\Theta$, which encompasses the trainable variables within the model. Additionally, $\lambda_1$ and $\lambda_2$ are hyperparameters that determine the respective strengths of the CL-based loss and the $L_2$ regularization term.

\section{Experiments}
\label{sec:exp}
To evaluate the effectiveness of our \model, we have designed a series of experiments to address the following research questions:
\begin{itemize}[leftmargin=*]

\item \textbf{RQ1}: How does the performance of our \model compare to a diverse range of state-of-the-art recommendation systems?

\item \textbf{RQ2}: What distinct contributions do the key components of our \model offer to the overall performance? Additionally, how does the model's performance adapt and respond to variations in hyperparameter settings?

\item \textbf{RQ3}: How does our proposed \model demonstrate its effectiveness in overcoming the obstacles of data sparsity and noise?

\item \textbf{RQ4}: To what degree does our proposed \model model provide a high level of interpretability for recommendation, facilitating a thorough comprehension of its decision-making process?

\end{itemize}

\begin{table}[t]
  \caption{Statistics of the experimental datasets.}
  \vspace{-0.1in}
  \label{tab:dataset}
  \centering
  \begin{tabular}{lccc}
    \hline
    Statistics & Last-FM & MIND & Alibaba-iFashion \\
    \hline
    $\#$ Users & 23,566 & 100,000 & 114,737 \\
    $\#$ Items & 48,123 & 30,577 & 30,040 \\
    $\#$ Interactions & 3,034,796 & 2,975,319 & 1,781,093 \\
    $\#$ Density & 2.7$e$-3 & 9.7$e$-4 & 5.2$e$-4 \\
    \hline
    Knowledge Graph \\
    \hline
    $\#$ Entities & 58,266 & 24,733 & 59,156 \\
    $\#$ Relations & 9 & 512 & 51 \\
    $\#$ Triplets & 464,567 & 148,568 & 279,155 \\
    \hline
  \end{tabular}
  \vspace{-0.1in}
\end{table}

\subsection{Experimental Settings}
\subsubsection{\bf Dataset.}
To ensure a comprehensive and diverse evaluation, we have incorporated three distinct public datasets that represent different real-life scenarios: Last-FM (music), MIND (news), and Alibaba-iFashion (e-commerce). To preprocess the data, we have applied the 10-core technique, filtering out users and items with occurrence counts below 10. For the Last-FM dataset, we have employed a mapping approach to associate the items with Freebase entities and extract knowledge triplets, following methodologies inspired by~\cite{wang2019kgat} and~\cite{zhao2019kb4rec}. In the case of the MIND dataset, we have followed the practices outlined in~\cite{tian2021joint} to collect the knowledge graph (KG) from Wikidata, focusing on representative entities within the original data. As for the Alibaba-iFashion dataset, we have manually constructed the KG, utilizing the category information as valuable knowledge~\cite{wang2021learning}. Detailed statistics for the three datasets and their corresponding KGs can be found in Table~\ref{tab:dataset}.

\subsubsection{\bf Evaluation Protocols.}
To avoid bias from negative sampling in evaluation~\cite{krichene2020sampled}, we report performance metrics under the full-rank setting, as done in the research works~\cite{wang2019kgat, wang2021learning,ren2023representation}. We utilize Recall@N and NDCG@N as top-N recommendation metrics, with N=20, a commonly used value~\cite{he2017neural, wang2019kgat}.

\subsubsection{\bf Compared Baseline Methods.}
For a comprehensive evaluation, we thoroughly compare our \model with a diverse set of baselines derived from different research streams.

\noindent\textbf{Collaborative Filtering Methods.}

\begin{itemize}[leftmargin=*]
    \item \textbf{BPR}~\cite{rendle2012bpr}: This method effectively utilizes pairwise ranking loss derived from implicit feedback for matrix factorization.
    \item \textbf{NeuMF}~\cite{he2017neural}: It incorporates MLP into matrix factorization and learns enriched user and item representations while capturing the feature interactions between them.
    \item \textbf{GC-MC}~\cite{berg2017graph}: By proposing a graph auto-encoder, GC-MC predicts unknown ratings by exploiting the underlying graph structure.
    \item \textbf{LightGCN}~\cite{he2020lightgcn}: Conducting an in-depth analysis of modules within standard GCN for collaborative data, LightGCN proposes a simplified GCN model tailored specifically for graph CF task.
    \item \textbf{SGL}~\cite{wu2021self}: SGL introduces data augmentation techniques such as random walk and feature dropout to generate multiple views.
\end{itemize}

\noindent \textbf{Embedding-based Knowledge-aware Recommenders.}
\begin{itemize}[leftmargin=*]
    \item \textbf{CKE}~\cite{zhang2016collaborative}: By integrating collaborative filtering and KG embeddings, CKE empowers the recommendation system with a deeper understanding of item relationships.
    \item \textbf{KTUP}~\cite{cao2019unifying}: This approach enables mutual complementation between collaborative filtering and knowledge graph signals, allowing for a more comprehensive recommendation process.
\end{itemize}

\noindent \textbf{GNN-based KG-enhanced Recommenders.}
\begin{itemize}[leftmargin=*]
    \item \textbf{KGNN-LS}~\cite{wang2019knowledgeb}: KGNN-LS considers user preferences towards different knowledge triplets in graph convolution. It introduces label-smoothing as regularization to encourage similar user preference weights between closely related items in the KG.
    \item \textbf{KGCN}~\cite{wang2019knowledge}: It aggregates knowledge for item representations by incorporating high-order information using GNNs.
    \item \textbf{KGAT}~\cite{wang2019kgat}: It introduces the concept of collaborative KG to apply attentive aggregation on the joint user-item-entity graph.
    \item \textbf{KGIN}~\cite{wang2021learning}: It models user intents for relations and employs relational path-aware aggregation to capture rich information from the composite knowledge graph.
\end{itemize}

\noindent \textbf{Self-Supervised Knowledge-aware Recommenders.}
\begin{itemize}[leftmargin=*]
    \item \textbf{MCCLK}~\cite{zou2022multi}: It employs contrastive learning in a hierarchical manner. It aims to mine useful structural information from the user-item-entity graph and its subgraphs. 
    \item \textbf{KGCL}~\cite{yang2022knowledge}: By leveraging self-supervised learning, KGCL effectively incorporates KG information while addressing noise and improving recommendation accuracy.
\end{itemize}

\subsection{RQ1: Overall Performance Comparison}

\begin{table}[t]
    \centering
    \caption{Performance comparison on Last-FM, MIND, Alibaba-iFashion datasets in terms of \textit{Recall@20} and \textit{NDCG@20}.}
    \resizebox{\linewidth}{!}{
    \begin{tabular}{c|c c|c c|c c}
        \hline
        \multirow{2}{*}{Model} & \multicolumn{2}{c|}{Last-FM} &  \multicolumn{2}{c|}{MIND} &  \multicolumn{2}{c}{Alibaba-iFashion} \\
        \cline{2-3}
        \cline{4-5}
        \cline{6-7}
        & Recall & NDCG & Recall & NDCG & Recall & NDCG \\
        \hline
        \hline
        BPR & 0.0690 & 0.0585 & 0.0384 & 0.0253 & 0.0822 & 0.0501 \\
        NeuMF & 0.0699 & 0.0615 & 0.0308 & 0.0237 & 0.0506 & 0.0276 \\
        GC-MC & 0.0709 & 0.0631 & 0.0386 & 0.0261 & 0.0845 & 0.0502 \\
        LightGCN & 0.0738 & 0.0647 & 0.0419 & 0.0253 & 0.1058 & 0.0652 \\
        SGL & 0.0879 & 0.0775 & \underline{0.0429} & 0.0275 & 0.1141 & 0.0713 \\
        CKE & 0.0845 & 0.0718 & 0.0387 & 0.0247 & 0.0835 & 0.0512 \\
        KTUP & 0.0865 & 0.0671 & 0.0362 & 0.0302 & 0.0976 & 0.0634 \\
        KGNN-LS & 0.0881 & 0.0690 & 0.0395 & 0.0302 & 0.0983 & 0.0633 \\
        KGCN & 0.0879 & 0.0694 & 0.0396 & \underline{0.0302} & 0.0983 & 0.0633 \\
        KGAT & 0.0870 & 0.0743 & 0.0340 & 0.0287 & 0.0957 & 0.0577 \\
        KGIN & 0.0900 & \underline{0.0779} & 0.0357 & 0.0225 & 0.1144 & \underline{0.0723} \\
        MCCLK & 0.0671 & 0.0603 & 0.0327 & 0.0194 & 0.1089 & 0.0707 \\
        KGCL & \underline{0.0905} & 0.0769 & 0.0399 & 0.0247 & \underline{0.1146} & 0.0719 \\
        \hline
        \model & \textbf{0.0980} & \textbf{0.0911} & \textbf{0.0615} & \textbf{0.0389} & \textbf{0.1234} & \textbf{0.0773} \\
        \hline
    \end{tabular}}
    \label{tab:performance}
    \vspace{-0.15in}
\end{table}

We have evaluated the overall performance of all the methods, and the results are summarized in Table~\ref{tab:performance}. Based on the findings, we have made the following observations:
\begin{itemize}[leftmargin=*]

    \item The performance evaluation of all methods consistently demonstrates that our proposed \model outperforms all baseline approaches. This highlights the effectiveness of our \model in enhancing recommendations with task-relevant KG signals. Specifically, our carefully designed graph diffusion model serves as a powerful graph generator, producing knowledge graphs that incorporate task-specific entity relationships. This enriched knowledge graph enhances the effectiveness of data augmentation, resulting in improved recommendation accuracy. \\\vspace{-0.1in}

    \item The performance evaluation clearly demonstrates the superiority of knowledge-aware recommenders that incorporate knowledge graph information compared to traditional approaches like BPR and NeuMF. This highlights the valuable role of knowledge graphs in addressing the sparsity issue inherent in collaborative filtering. The noticeable performance gap between our \model and other knowledge-aware models, such as KGAT, KGIN, and KGCL, suggests that knowledge graphs often contain irrelevant relations that can negatively impact recommendation quality. \\\vspace{-0.1in}

    \item The comparative performance of KGCL highlights the effectiveness of incorporating KG-based item semantic relatedness and leveraging self-supervised signals to explicitly address the interaction sparsity issue. KGCL focuses on augmenting the user-item interaction matrix with KG guidance, while our \model takes a different approach by utilizing a task-related knowledge graph generated through our designed KG diffusion model.
    
\end{itemize}

\subsection{RQ2: Ablation Study}
\subsubsection{\bf Key Module Ablation.}
This study aims to evaluate the effectiveness of the key modules incorporated in our proposed \model. To establish a comparative analysis with the original method, we have developed three distinct model variants, which are outlined:
\begin{itemize}[leftmargin=*]
\item "w/o CL": This variant involves the removal of the KG-enhanced data augmentation module in recommendation.
\item "w/o DM": We replace our diffusion model with variational graph autoencoder, which is a widely-used generative model.
\item "w/o CKGC": This variant excludes the collaborative knowledge graph convolution from the KG diffusion model optimization.
\end{itemize}

The ablation study results, as presented in Table~\ref{tab:exp_ablation}, yield important insights, leading to the following key conclusions: i) Removal of KG-enhanced contrastive learning results in significant performance degradation across all cases. This finding validates the effectiveness of incorporating additional self-supervised signals using the knowledge graph. ii) Ablation of the knowledge graph diffusion model component demonstrates its crucial role in improving the performance of our \model. In all cases, the inclusion of our designed diffusion model contributes to better results, affirming the effectiveness of capturing task-relevant relations through the diffusion process. Notably, the larger performance drop observed in Last-FM and MIND datasets suggests a higher level of noise present in their respective knowledge graphs. iii) The absence of the collaborative knowledge graph convolution module leads to performance degradation across all cases. This underscores the significance of collaborative knowledge graph convolution in our \model, as it facilitates the integration of user collaborative knowledge into the training of the diffusion model for recommendation.

\subsubsection{\bf Sensitivity to Key Hyperparameters.}
\begin{figure}[]
    \centering
    \subfigure[Hyperparameter Analysis on Data Augmentation]{
    \label{fig:figure_exp_hyper_cl}
    \includegraphics[width=0.49 \linewidth]{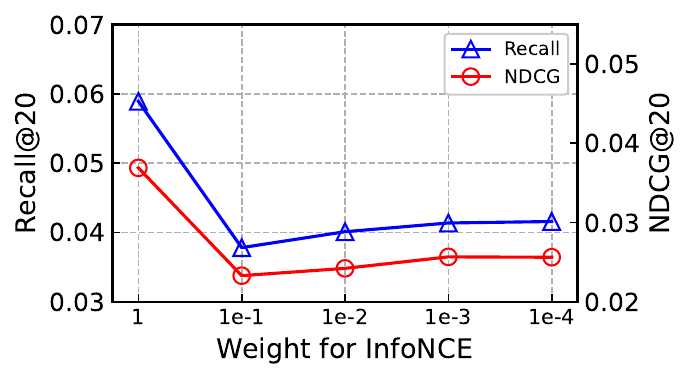}
    \includegraphics[width=0.49 \linewidth]{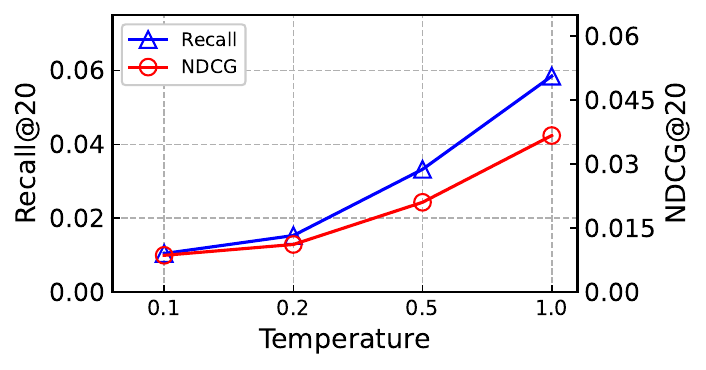}}
    \subfigure[Hyperparameter Analysis on Diffusion Model]{
    \label{fig:figure_exp_hyper_dm}
    \includegraphics[width=0.49 \linewidth]{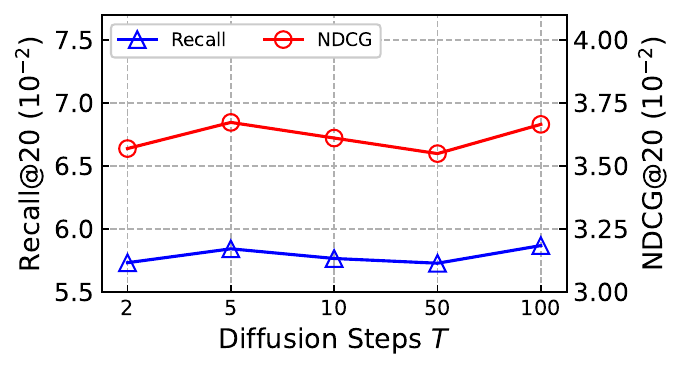}
    \includegraphics[width=0.49 \linewidth]{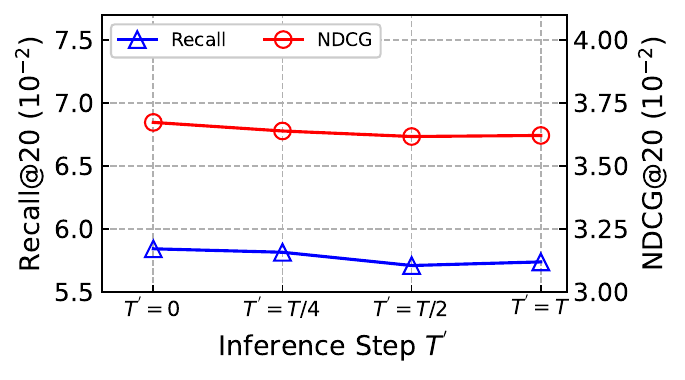}}
    \vspace{-0.2in}
    \caption{Hyperparameter Analysis on MIND Dataset.}
    \vspace{-0.15in}
    \label{fig:figure_exp_hyper}
\end{figure}

In this study, we focus on examining the effects of different hyperparameters on our method. Specifically, we conduct a thorough analysis of hyperparameters in both the data augmentation and knowledge graph diffusion modules. To present our findings, we report the corresponding results on the MIND dataset, as demonstrated in Figure~\ref{fig:figure_exp_hyper}.

We thoroughly analyzed hyperparameters for our \model, specifically focusing on $\lambda_1$ (InfoNCE loss weight) and $\tau$ (softmax temperature). Figure~\ref{fig:figure_exp_hyper_cl} showcased the best performance with $\lambda_1 = 1$ and $\tau = 1$, emphasizing the significance of CL. Additionally, in the knowledge graph diffusion model, Figure~\ref{fig:figure_exp_hyper_dm} demonstrated minimal accuracy impact when increasing diffusion steps due to low noise levels. We selected $\textit{T} = 5$ to balance performance and computation. Notably, the best performance was achieved with $\textit{T}^{'} = 0$ to avoid excessive corruption of the original KG.

\begin{table}[t]
    \centering
    \caption{Ablation study on key components of \model.}
    \vspace{-0.12in}
    \resizebox{\linewidth}{!}{
    \begin{tabular}{l|c c|c c|c c}
        \hline
        \multirow{2}{*}{Ablation Settings} & \multicolumn{2}{|c|}{Last-FM} &  \multicolumn{2}{|c|}{MIND} &  \multicolumn{2}{|c}{Alibaba-iFashion} \\
        \cline{2-7}
        & Recall & NDCG & Recall & NDCG & Recall & NDCG \\
        \hline
        \model & \textbf{0.0980} & \textbf{0.0911} & \textbf{0.0615} & \textbf{0.0389} & \textbf{0.1234} & \textbf{0.0773} \\
        \hline
        \model w/o CL & 0.0737 & 0.0638 & 0.0398 & 0.0246 & 0.1126 & 0.0700 \\
        \model w/o DM & 0.0961 & 0.0891 & 0.0573 & 0.0358 & 0.1218 & 0.0767 \\
        \model w/o CKGC & 0.0976 & 0.0905 & 0.0605 & 0.0378 & 0.1228 & 0.0770 \\
        \hline
    \end{tabular}
    }    
    \vspace{-0.2in}
    \label{tab:exp_ablation}
\end{table}

\subsection{RQ3: Further Investigation on \model}

\begin{figure}[]
    \centering
    \subfigure[Performance \textit{w.r.t.} cold-start user groups]{
    \label{fig:figure_exp_sp_user}
    \includegraphics[width=0.49 \linewidth]{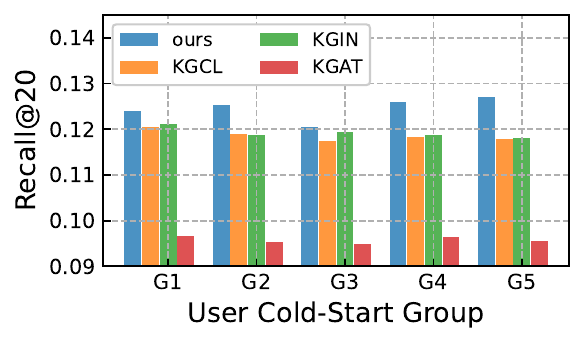}
    \includegraphics[width=0.49 \linewidth]{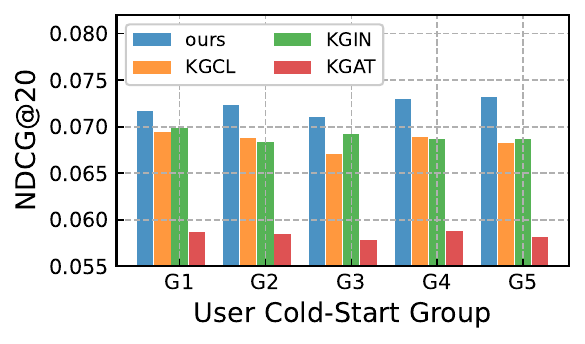}}
    \subfigure[Performance \textit{w.r.t.} sparse item groups]{
    \label{fig:figure_exp_sp_item}
    \includegraphics[width=0.49 \linewidth]{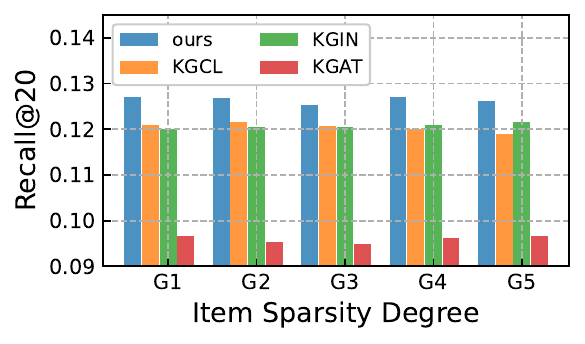}
    \includegraphics[width=0.49 \linewidth]{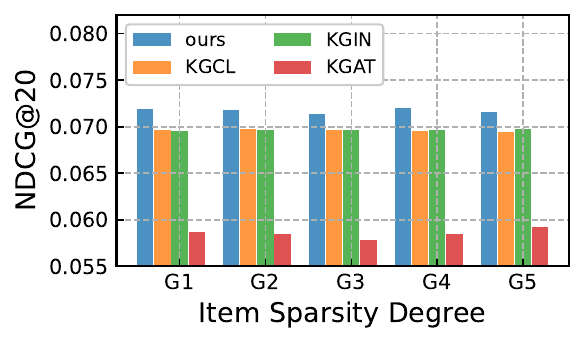}}
    \vspace{-0.15in}
    \caption{Performance \textit{w.r.t} different data sparsity degrees.}
    \vspace{-0.1in}
    \label{fig:figure_exp_sparsity}
\end{figure}

\noindent\textbf{Sparse User Interaction Data.}
In order to assess the performance of our \model in handling sparse data, we conducted an evaluation on both users and items. For users, we divided them into five groups, each containing an equal number of users. The interaction density within these groups gradually increased from Group 1 to Group 5, representing varying levels of sparsity. A similar approach was employed for processing the items. The test results from this evaluation are presented in Figure~\ref{fig:figure_exp_sparsity}. \\\vspace{-0.1in}

\begin{figure}[t]
    \centering
    \subfigure[Relative Recall]{
    \includegraphics[width=0.48 \linewidth]{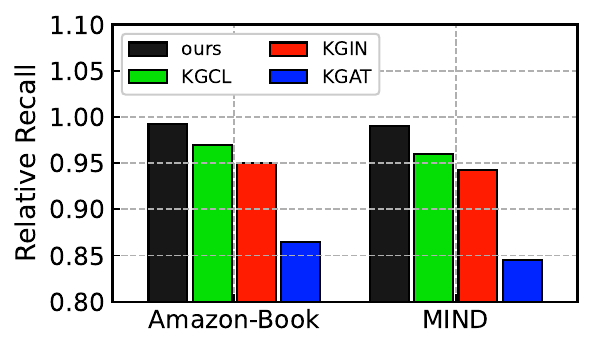}}
    \subfigure[Relative NDCG]{
    \includegraphics[width=0.48 \linewidth]{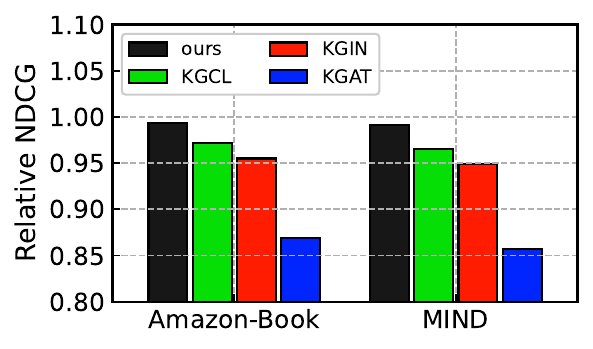}}
    \vspace{-0.15in}
    \caption{Performance in alleviating KG noise.}
    \vspace{-0.1in}
    \label{fig:figure_exp_noise}
\end{figure}

\noindent\textbf{Knowledge Graph Noise.}
To assess our \model's ability to filter out irrelevant relations from the KG, we injected noisy triplets into the data and compared its performance with other knowledge-aware recommender systems. Specifically, we randomly added 10\% noisy triplets to the existing KG while keeping the test set unchanged, simulating a scenario with a large number of topic-irrelevant relations. The test results can be found in Fig.~\ref{fig:figure_exp_noise}.

\begin{itemize}[leftmargin=*]
    \item The evaluation of sparse data recommendation clearly demonstrates the superior performance of our \model compared to KGCL. This notable improvement serves as strong evidence of the effectiveness of our KG-enhanced data augmentation. It effectively tackles the challenge posed by task-irrelevant relations within the KG, which have the potential to mislead the encoding of user preferences in the recommendation process. \\\vspace{-0.1in}

    \item In the recommendation scenario with long-tail item distributions, our \model significantly improves recommendation performance for such items. This highlights its effectiveness in mitigating popularity bias, as other baseline methods tend to neglect less popular items. Additionally, our \model outperforms competitive KG-aware recommendation systems like KGAT and KGIN. This suggests that blindly incorporating all KG information into collaborative filtering may introduce noise from irrelevant item relations and fail to alleviate popularity bias effectively. \\\vspace{-0.1in}
    
    \item Among the various knowledge-aware recommendation models, our \model consistently achieves the highest performance. This can be attributed to the task-specific knowledge graph generated by the diffusion model. Notably, \model demonstrates the most effective noise alleviation, as evidenced by the lowest average performance decrease in the presence of KG noise, as depicted in Figure~\ref{fig:figure_exp_noise}. This serves as compelling evidence of the remarkable ability of our \model to discover relevant information from a noisy KG, effectively supporting user preference modeling.
    
\end{itemize}

\vspace{-0.1in}
\subsection{RQ4: Case Study}

\begin{figure}[t]
    \centering
    \includegraphics[width=1 \linewidth]{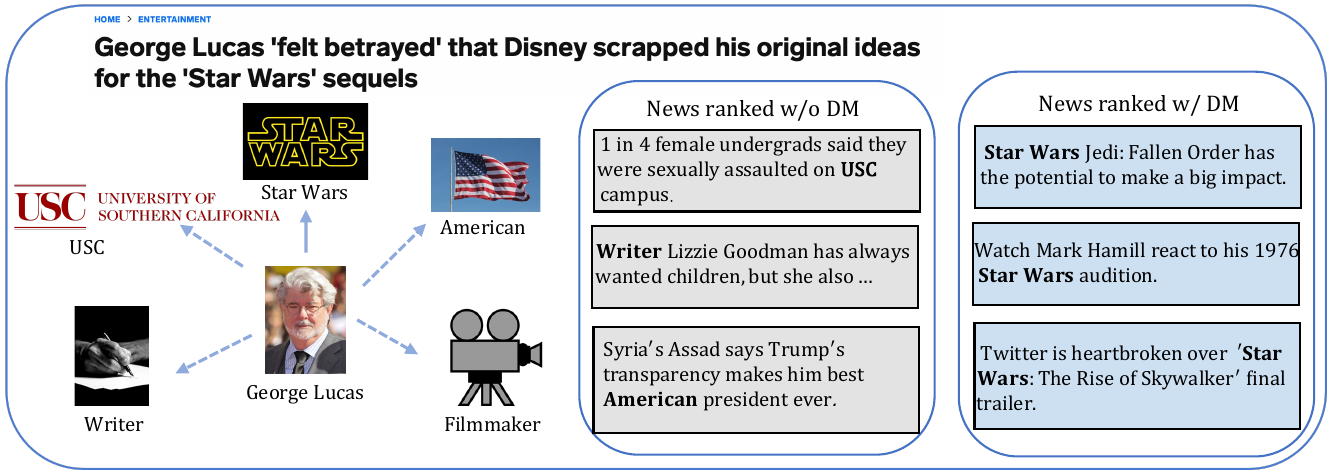}
    \vspace{-0.15in}
    \caption{Relevant News w/ and w/o KG diffusion.}
    \vspace{-0.15in}
    \label{fig:figure_case_study}
\end{figure}

We performed a case study on news recommendation, comparing the results with and without our knowledge graph diffusion model. The findings are shown in Figure~\ref{fig:figure_case_study}, highlighting the impact of our KG diffusion on recommendation accuracy. We examine the assessment of the Star Wars sequel by renowned filmmaker George Lucas and its relevance to the provided KG information. The KG includes entities such as "American," "USC," "writer," and "filmmaker," which are unrelated to the news at hand. This noise in the KG can introduce bias and misguide user representation. Without the knowledge graph diffusion model, the model ranks unrelated news articles covering topics such as "USC," "Lizzie Goodman" (a writer), and "Syria." However, with the integration of KG diffusion paradigm, our \model effectively filters out irrelevant KG information, resulting in more pertinent news articles. These articles include a discussion on a Star Wars video game, an actor's involvement in the Star Wars film, and social media commentary on the Star Wars movie. By accurately leveraging and filtering KG information, our model demonstrates improved performance in recommendation tasks, illustrating its effectiveness in enhancing relevance and mitigating the impact of irrelevant information in the KG.
\section{Related Work}
\subsection{Knowledge-aware Recommender Systems}
Existing knowledge-aware recommendation methods can be categorized into embedding-based, path-based, and GNN-based approaches. GNN-based methods, such as KGCN \cite{wang2019knowledge}, KGAT \cite{wang2019kgat}, and KGIN \cite{wang2021learning}, combine the strengths of both paradigms and effectively extract valuable information from the knowledge graph. KGCN utilizes a fixed number of neighbors for item representation aggregation, while KGAT employs Graph Attention Networks (GATs) to assign weights based on the importance of knowledge neighbors. KGIN incorporates user preferences and relational embeddings in the aggregation layer. These GNN-based methods enhance recommendation systems by leveraging the power of GNNs and the rich information in the knowledge graph \cite{wang2019knowledgeb, wang2019knowledge, wang2019kgat, wang2021learning}.

\subsection{Data Augmentation for Recommendation}
Data augmentation techniques, combined with self-supervised learning (SSL), have emerged as a promising approach to enhance recommendation systems. By leveraging additional supervision signals extracted from raw data, SSL-based data augmentation methods can address data sparsity and improve recommendation performance~\cite{wu2021self,chen2023heterogeneous}. Contrastive learning-based data augmentation methods, such as those proposed in~\cite{wu2021self,2023multi}, generate augmented views of user or item representations. By training models to differentiate between positive and negative pairs~\cite{li2023multi,yang2023debiased}, these methods effectively addresses data sparsity and enhances recommendation performance through self-supervised learning. Additionally, inspired by natural language processing tasks like BERT, masking and reconstruction augmentation techniques involve masking or hiding certain items or parts of user-item interactions and training the model to predict the missing elements. This process forces the model to learn contextual relationships in the recommendation process~\cite{sun2019bert4rec,ren2023distillation}. By incorporating SSL-based data augmentation techniques into recommendation systems, models can address data sparsity, capture complex patterns, and improve the generalization ability of recommender systems.

\subsection{Diffusion Probabilistic Models}
Diffusion probabilistic models have gained considerable attention and showcased great potential in a range of fields, spanning computer vision and natural language processing. In the context of vision, diffusion models have been particularly effective in tasks such as image generation~\cite{gu2022vector,ho2022cascaded} and inpainting~\cite{lugmayr2022repaint}. In the context of text generation, a generative model is trained to recover the original text from the perturbed data~\cite{li2022diffusion,gong2022diffuseq}. In addition, diffusion models have also found application in diverse domains, including graph learning for the purpose of graph generation. For example, GraphGDP~\cite{huang2022graphgdp} proposes a continuous-time generative diffusion process for permutation invariant graph generation. Digress~\cite{vignac2023digress} employs a discrete denoising diffusion model that utilizes a graph transformer network to iteratively modify graphs with noise, resulting in the generation of graphs. Recently, diffusion probabilistic models have also been explored in the realm of recommendation~\cite{wang2023diffusion}.

\section{Conclusion}
\label{sec:conclusion}
This research introduces \model, a novel recommendation model that leverages task-specific item knowledge to enhance the collaborative filtering paradigm. The framework proposes a unique methodology for extracting high-quality signals from noisy knowledge graphs. By seamlessly integrating a generative diffusion model with a knowledge graph learning framework tailored for knowledge-aware recommender systems, the model effectively aligns the semantic aspects of knowledge-enhanced items with collaborative relation modeling, resulting in precise recommendations. Through extensive evaluations on diverse benchmark datasets, our proposed \model framework demonstrates significant performance improvements compared to various baseline models. Furthermore, our approach effectively addresses the challenge of noisy data, which is known to impede the accuracy of recommender systems.

\clearpage
\balance
\bibliographystyle{ACM-Reference-Format}
\bibliography{main.bbl}

\clearpage
\appendix

\end{document}